\documentclass[aps,twocolumn,showpacs]{revtex4}

\usepackage{graphicx}
\usepackage{amsfonts}

\def\be{\begin{equation}}
\def\ee{\end{equation}}
\def\bea{\begin{eqnarray}}
\def\eea{\end{eqnarray}}

\def\1{\'{\i}}
\def\R{{\mathbb R}}

\def\cte{a}
\def\cteb{b}

\def\jj{j}
\def\mm{m}
\def\nn{n}
\def\cc{C}

\def\newi{\hat I}
\def\newcas{ {\hat {\bf I}} }

\begin{document}

\title{Generalized rotational Hamiltonians from non-linear angular momentum
algebras}\vskip2cm
\author{A. Ballesteros$^{a)}$}
\author{O. Civitarese$^{b)}$}
\author{F.J. Herranz$^{a)}$}
\author{M. Reboiro$^{b),c)}$}
\affiliation{{\small\it $^{a)}$Departamento de F\'{\i}sica,
Universidad de Burgos,} {\small\it Pza.~Misael Ba\~nuelos, E-09001
Burgos, Spain}} \affiliation{{\small\it $^{b)}$Department of
Physics, University of La Plata,} {\small\it c.c.~67 1900, La
Plata, Argentina}} \affiliation{{\small\it $^{c)}$Faculty of
Engineering, University of Lomas de Zamora} {\small\it C. C. Km 2,
(1836) Lavallol, Argentina} }
\date{\today}

\begin{abstract}
Higgs algebras are used to construct rotational Hamiltonians. The
correspondence between the spectrum of a triaxial rotor and the
spectrum of a cubic Higgs algebra is demonstrated. It is shown
that a suitable choice of the parameters of the polynomial algebra
allows for a precise identification of rotational properties. The
harmonic limit is obtained by a contraction of the algebra,
leading to a linear symmetry.
\end{abstract}



\maketitle

\section{Introduction}

Mathematical aspects of the Higgs algebras have been studied
intensively in the past \cite{r2,r1,r3}. Generally speaking, in
the framework  of non-linear algebras, the commutation relations
between the generators are expressed as a linear combination of
power terms of them, this is the case of the $sl^{(3)}(2,\R)$
algebra, where the commutator of the ladder operators contains up
to cubic powers of $J_3$. In spite of their relative involved
structures, the fact that the algebras contain as limiting cases
the more familiar structures of the groups associated to rotations
and vibrations turns out to be a particularly appealing feature
for applications to the  group classification of nuclear
Hamiltonians \cite{ref4,ref5}. Therefore, they may be useful tools
to classify observables. The search for physical inspired
Hamiltonians, which may display definite features about polynomial
algebras is still open. Concrete applications of the formalism
have been explored more recently in the context of schematic
models \cite{ref6}, to complement previous mathematical efforts,
like the studies of \cite{ref6b,ref6c,ref7}. Among the already
studied applications of the concept of non-linear algebras, the
results of \cite{ref7} can be taken as definite motivations for
our present effort.

In this work we continue with the study of applications of Higgs
algebras, by constructing a Hamiltonian which reproduces the
behavior of an asymmetric rotor and its vibrational limit
\cite{ref4}. We shall show that, while the finite dimensional
representations of non-linear algebras are suitable for the
description of rotational-like structures, the infinite
dimensional representations of a Higgs algebra contain
vibrational-like structures.

The paper is organized as follows. In Section II we present the
essentials of the formalism, and introduce both the infinite- and
finite-dimensional representations of  $sl^{(3)}(2,\R)$. In
Section III.A, we present the basic concepts which we have
developed in the construction of a triaxial rotor Hamiltonian. In
Section III.B we study the vibrational limit of this Hamiltonian
by contracting a particular, $sl^{(3)}(2,\R)$, Higgs algebra.
Numerical results are presented in Section IV. Conclusions are
drawn in Section V.

\section{Polynomial angular momentum algebras}

To start with we shall consider polynomial angular momentum
algebras up to the third power in $J_3$ \cite{r2,r3}. The
commutation relations of the $sl^{(3)}(2,\R)$ algebra are defined
by \cite{r2,r1,r3}

\be
\begin{array}{l}
[J_3,J_\pm]=\pm J_\pm, \\[2pt]
[J_+,J_-]=\alpha J_3^3+\beta J_3^2+\gamma J_3 +\delta ,
\end{array}
\label{sl3}
\ee
where the parameters $\alpha,\beta,\gamma$, and $\delta$ take arbitrary real values.
The corresponding Casimir operator reads
\bea
&&{\cal C}=J_+ J_- +\frac{\alpha}{4}
J_3^4+\left(\frac{\beta}{3} -\frac{\alpha}{2}  \right)
J_3^3\nonumber\\
&&\qquad +\left( \frac{\alpha}{4}-\frac{\beta}{2}
+\frac{\gamma}{2}  \right)J_3^2 +\left( \frac{\beta}{6}
-\frac{\gamma}{2}+\delta  \right)J_3 . \label{ab} \eea


Notice that $sl^{(3)}(2,\R)$ comprises, as particular cases,
several well-known algebras. In the linear case ($\alpha=\beta=0$)
we recover the following Lie algebras: i) $sl(2,\R)$ if $\delta=0$
and $\gamma$ takes any non-zero real value; (ii) $gl(2)$ if both
$\gamma$ and $\delta$  take any non-zero real value; (iii) the
oscillator algebra $h_4$ if $\gamma=0$ and $\delta$ takes any
non-zero real value; and (iv) the Poincar\'e algebra  for
$\gamma=\delta=0$ \cite{r2}.   In all these cases, when a
parameter (either $\gamma$ or $\delta$) is different from zero, it
can be  reduced to $\pm 1$ by re-scaling the Lie generators. On
the other hand, the quadratic algebra $sl^{(2)}(2,\R)$ is obtained
when $\alpha=0$ and $\beta\neq 0$.  In this case, the linear term
$\gamma J_3$ can be eliminated by making use of the automorphism
$J_3\rightarrow J_3 - \psi$, where $\psi$ is a constant that has
to be fitted. We recall that $sl^{(2)}(2,\R)$  is, for appropriate
values of the constants $\beta,\gamma$ and $\delta$, the dynamical
algebra for the Dicke and Da Providencia-Sch\"utte models (see
\cite{tmp} and references therein). We also remark that in the
cubic case given by $\alpha\neq 0$, the same kind of automorphism
allows us to eliminate the quadratic term $\beta J_3^2$. Moreover,
the case with $\beta=\delta=0$  corresponds to the well-known
Higgs algebra \cite{r2}. This cubic algebra was found in the study
of the Kepler potential on a two-dimensional space with constant
curvature and has motivated further developments of the
representation theory of polynomial algebras \cite{r1,r3}.
Finally, note that from the complementary viewpoint, the structure
constants $\alpha,\beta,\gamma,\delta$ in (\ref{sl3}) can be
interpreted as deformation parameters that generate a wide class
of structures by taking the Poincar\'e algebra  (with
$\alpha=\beta=\gamma=\delta=0$) as the origin of the space of
deformations.


As far as the finite-dimensional representations of
$sl^{(3)}(2,\R)$  is concerned, these are defined on the
$(2\jj+1)$-dimensional basis spanned by the usual vectors
$|\jj\mm\rangle$ with $\jj=0,\frac 12,1,\frac 32,\dots$ and
$\mm=-\jj,\dots,0,\dots,\jj$, like, e.g.~\cite{r2} \bea
&&J_3|\jj\mm\rangle=\left(\frac{\mm}{q}+c \right)|\jj\mm\rangle,\nonumber\\
&&J_+|\jj\mm\rangle=f(\mm)|\jj\mm+q\rangle,\nonumber \\
&&J_-|\jj\mm\rangle=g(\mm)|\jj\mm-q\rangle, \label{ac} \eea where
$q$ is a positive integer number and $c$ is a real number. For a
given $\jj$ the product of functions $f(\mm)$ and $g(\mm)$ fulfils

\be
\begin{array}{l}
f(\mm-q)g(\mm)-f(\mm)g(\mm+q)  \\[4pt]
\displaystyle{= \alpha\left(\frac{\mm}{q}+c \right)^3+\beta \left(\frac{\mm}{q}+c \right)^2
+\gamma \left(\frac{\mm}{q}+c \right)+\delta,}
\end{array}
\label{ad}
\ee
for $\mm=-\jj,\dots,\jj$, and the functions $f$ and $g$ also
obey the constraints
\be
\begin{array}{l}
f(\jj)=f(\jj-1)=\dots =f(\jj-q+1)=0, \\
g(-\jj)=g(-\jj+1)=\dots =g(-\jj+q-1)=0.
\end{array}
\label{ae1}
\ee
Some other finite-dimensional representations can
be found in~\cite{r1,r3}.

Infinite-dimensional representations of $sl^{(3)}(2,\R)$ can be
built on the Hilbert space of number states spanned by the
set $\{ |\nn\rangle \}$, namely:
\bea
&&J_3|\nn\rangle=\left(\frac{\nn}{q}+c \right)|\nn\rangle, \nonumber\\
&&J_+|\nn\rangle={\cal F}(\nn)|\nn+q\rangle, \nonumber \\
&&J_-|\nn\rangle={\cal G}(\nn)|\nn-q\rangle.
\label{ag}
\eea
The functions ${\cal F}(\nn)$ and ${\cal G}(\nn)$ of Eq.
(\ref{ag}) satisfy the conditions

\be
\begin{array}{l}
{\cal F}(\nn-q){\cal G}(\nn)-{\cal F}(\nn){\cal G}(\nn+q) \nonumber\\[4pt]
 \displaystyle{\quad =  \alpha\left(\frac{\nn}{q}+c
\right)^3+\beta \left(\frac{\nn}{q}+c \right)^2+\gamma \left(\frac{\nn}{q}+c
\right)+\delta}, \nonumber\\[8pt]
{\cal G}(0)={\cal G}(1)\dots ={\cal G}(q-1)=0,
\end{array}
\label{ah}
\ee
which can be compared with the conditions given by
Eqs.(\ref{ac})-(\ref{ae1}).

The expressions appearing in (\ref{ac}) suggest that the
generators $J_\pm$ play the role of ladder operators with step $q$
acting on the state $|\jj\mm\rangle$. Likewise, the results shown
in equations  (\ref{ag}) relate $J_{\pm}$ to ladder operators of a
Lie algebra to power $q$ \footnote{Generally speaking a boson-like
of the form $J_+ \propto {a^\dagger}^q$, and $J_- \propto
{a}^q$.}. In view of the correspondence of Eqs.
(\ref{ac})-(\ref{ae1}) and Eqs. (\ref{ag})-(\ref{ah}) we stress
the fact that the structure of $sl^{(3)}(2,\R)$ allows for a
generalization of the definition of the usual angular momentum
operators in the rotational limit (finite-dimensional
representation) and also a generalization of the harmonic
oscillator in the vibrational limit (infinite-dimensional
representations). In the next section we shall advanced on these
concepts.

\section{Generalised Hamiltonians.}

If $sl^{(3)}(2,\R)$ is interpreted as a dynamical algebra, then
the most general Hamiltonian  based on such non-linear symmetry
takes the form \be {\cal H}=\sum_{i=0}^4 ~\cc_i J_3^i+ \cc_+ J_+ +
\cc_- J_-, \label{ca} \ee where the coefficients $\cc_i$ as well
as $\cc_\pm$ are arbitrary real numbers.

In order to obtain explicit models, let us introduce the following
finite-dimensional representation (\ref{ac})-(\ref{ad}) with
$q=2$. The functions $f(\mm)$ and $g(\mm)$ are, for this case

\be
\begin{array}{l}
f(\mm) = \cte \sqrt{(\jj+\mm+1)(\jj+\mm+2)(\jj-\mm)(\jj-\mm-1)} , \\[2pt]
g(\mm) =  \cte \sqrt{(\jj+\mm)(\jj+\mm-1)(\jj-\mm+1)(\jj-\mm+2)},
\end{array}
\label{cb} \ee where $\cte$ is a real constant. Notice that these
functions resemble the eigenvalues of the square of the linear
$sl(2,\R)$ ladder operators. Next, according to Eq. (\ref{ad}),
the deformation parameters $\alpha,\beta,\gamma$, and $\delta$ in
(\ref{sl3}) turn out to be

\be
\begin{array}{l}
\alpha=-64\cte^2,\qquad \beta=192\cte^2 c,\\[2pt]
\gamma=-8\cte^2\left(1+24 c^2- 2 j(j+1) \right),\\[2pt]
\delta=8\cte^2 c\left(1+8 c^2- 2 j(j+1) \right).
\end{array}
\label{cc}
\ee
In this representation, the eigenvalue of the
Casimir operator (\ref{ab}) reads

\begin{eqnarray}
\langle \jj\mm| {\cal C}|\jj\mm\rangle
& = &  \cte^2 \left(j(j+1) - 2 c - 4 c^2 \right) \nonumber \\
&   & \quad \times \left(j(j+1) - 2- 6 c - 4 c^2 \right).
\label{cd}
\end{eqnarray}
The non-vanishing matrix elements of the
Hamiltonian (\ref{ca}) are given by

\be
\begin{array}{l}
\displaystyle{  \langle \jj\mm| {\cal H}|\jj\mm\rangle=
\sum_{i=0}^4\cc_i {\left(\frac{\mm}{2}+c \right)  }^i} \nonumber \\
\displaystyle{\langle \jj\mm+2| {\cal H}|\jj\mm\rangle= \cc_+~f(m)
}, \nonumber \\
\displaystyle{\langle \jj\mm-2| {\cal
H}|\jj\mm\rangle= \cc_-~g(m) }.
\end{array}
\label{ce} \ee The above expressions demonstrate that
finite-dimensional representations of non-symmetric algebras are,
therefore, suitable for the treatment of rotational-like
structures. This property is a result of the explicit dependence
of the functions $f$ and $g$ upon $j$, which is acting as the
effective level of the representation since $f(m)$ and $g(m)$
vanish if $m=j,j-1$ and $m=-j,-j+1$, respectively.

Similarly, for  the case of
infinite-dimensional representations one has

\be
\begin{array}{l}
{\cal F}(\nn) = \cteb \sqrt{(\nn+1)(\nn+2)} , \\[2pt]
{\cal G}(\nn) =  \cteb \sqrt{\nn(\nn-1)},
\end{array}
\label{da}
\ee
provided that $q=2$, $\cteb$ is a real constant,
and the deformation parameters are

\be
\alpha=
\beta= 0,\quad \gamma= - 8 \cteb^2,\quad
\delta=2\cteb^2(4 c-1).
\label{db}
\ee
Hence the eigenvalue of the Casimir operator (\ref{ab}) is

\be
\langle n | {\cal C}| n\rangle=
2\cteb^2 c(1+ 2 c).
\label{dc}
\ee
For this case the non-vanishing matrix elements of
the Hamiltonian (\ref{ca}) are
\be
\begin{array}{l}
\displaystyle{  \langle  n | {\cal H}| n \rangle=
\sum_{i=0}^4  \cc_i \left(\frac {n}{2}+c \right)^i  },\\[6pt]
\displaystyle{\langle n+2| {\cal H}| n \rangle= \cc_+ {\cal F}(n) }, \\[2pt]
\displaystyle{\langle n-2| {\cal H}| n \rangle= \cc_- {\cal G}(n) },
\end{array}
\label{de} \ee to be compared with (\ref{ce}). Thus,
infinite-dimensional representations of non-linear algebras
exhibit vibrational-like structures, under a proper choice of
parameters, based on the explicit dependence of ${\cal F}$ and
${\cal G}$ upon $n$.

\subsection{Triaxial rotor}

The results of the previous subsection illustrate rather
explicitly the scope of the formal connection we would like to
establish between finite (infinite) representations of a
polynomial algebra and the generators of angular momentum
(vibrational quanta) and the corresponding Hamiltonians.

Let us start with the case of a triaxial rotor \cite{ref4}, whose
Hamiltonian is written

\be H_{\rm{rot}}=\sum_{i=1}^3 A_i I^2_i \label{ea} \ee where
\begin{eqnarray}
A_i = \frac{1}{2 {\cal I}_i}.
\end{eqnarray}
The operators $I_i$ are the components of the angular momentum
along the principal axes, and ${\cal I}_i$ are the components of
the tensor of inertia. The generators $I_i$, of the intrinsic
angular momentum components respect to the body-fixed frame,
satisfy the commutation relations $\left [ I_i,I_j \right ]= -{\rm
i }\varepsilon_{ijk} I_k$. We define the angular momentum rising
and lowering ladder operators

\begin{eqnarray}
I_\pm =  I_1 \pm {\rm  i} I_2 ,
\label{eb}
\end{eqnarray}
which obey the commutation rules

\be [I_3,I_\pm]=\mp I_\pm,\quad [I_+,I_-]=-2I_3. \label{ec} \ee
The Casimir invariant is given by \be {\bf I}^2=\sum_{i=1}^3
I_i^2=I_3^2+ \frac 12  (I_+ I_- + I_- I_+). \label{ed} \ee In this
way the Hamiltonian (\ref{ea}) can be written as
\begin{eqnarray}
H & = & \frac 1 2 (A_1+A_2) {\bf I}^2 + \frac 14
(A_1-A_2) (I_+^2+ I_-^2)\nonumber \\ & &
+ \frac 12 (A_1-A_2) \kappa I_3^2,
\label{hrotor}
\end{eqnarray}
where \be \kappa =\frac {2 A_3-A_1-A_2}{A_1-A_2}, \label{ef} \ee
is the factor which measures the asymmetry of the rotor.

In the following, we shall show that a correspondence between the
triaxial rotor case and a cubic Higgs algebra can indeed be
established by a proper choice of the parameters of the non-linear
symmetry.

Firstly, we adopt the following choice of parameters, in
Eq.(\ref{ab}),
\begin{eqnarray}
\alpha=-\frac{4}{I^2},~~ \beta=0,~~ \gamma=\frac {1}{2 I^2}(2
I^2+2I -1),~~ \delta=0, \nonumber \\ \label{choi}
\end{eqnarray}
for the non-linear symmetry, and taking $q=2$, $c=0$, and $a=\frac
1{4I}$ in Eq.(\ref{ad}) one obtains
\begin{eqnarray}
f(M)=a \sqrt{(J+M+1)(J+M+2)(J-M)(J-M-1)}&&   \nonumber \\
g(M)=a \sqrt{(J+M)(J+M-1)(J-M+1)(J-M+2)},&& \nonumber \\
\end{eqnarray}
where $J=I$ and $M=M_I$, being $I$ the total angular momentum and
$M_I$ its projection. The operators $J_\pm$ and $J_3$ are written

\begin{eqnarray}
J_\pm = \frac{1}{4 I} I_\pm^2, ~~~ J_3= -\frac 1 2 I_3.
\end{eqnarray}
The corresponding commutation relations are
\begin{eqnarray}
\left [ J_3, J_{\pm} \right ] & = & \pm J_{\pm},  \nonumber \\
\left [ J_+, J_- \right ]     & = & \frac {1} {2 I^2 } (2
I^2+2I-1) ~ J_3- \frac {4}{I^2} ~ J_3^3. \label{algparticular}
\end{eqnarray}
Secondly, the Hamiltonian of Eq. (\ref{hrotor}) is expressed in
terms of the non-linear symmetry as
\begin{eqnarray}
{\cal H } & = & \frac 1 2 (A_1+A_2) J (J+1)  + J (A_1-A_2) (J_+
+ J_-)\nonumber \\ & & + 2 (A_1-A_2) \kappa J_3^2.
\label{hrotorhiggs}
\end{eqnarray}
The matrix elements of the Hamiltonian of Eq. (\ref{hrotorhiggs}) are given by
\begin{equation}
\begin{array}{l}
\displaystyle{ \langle {\rm J~M}| {\cal H}|{\rm J M} \rangle =
\frac 12 (A_1+A_2) J(J+1)     +    \frac 12 (A_1-A_2)\kappa M^2  },\\[2pt]
\displaystyle{\langle {\rm J~M+2}| {\cal H}|{\rm J M} \rangle= J(A_1-A_2) f(M) }, \\[2pt]
\displaystyle{\langle {\rm J~M-2}| {\cal H}|{\rm J M} \rangle= J(A_1-A_2) g(M) }.
\end{array}
\label{ej}
\end{equation}
Both Hamiltonians, the one of Eq. (\ref{hrotor}) and the one of
Eq.(\ref{hrotorhiggs}), have the same matrix elements, and in
consequence the same spectrum and the same eigenfunctions.

The above example shows that the standard angular momentum
operators of the triaxial rotor can be understood as the mapped
version of the generators of a particular non-linear symmetry.
That is to say that the triaxial rotor is a particular case of a
finite-dimensional representation of $sl^{(3)}(2,\R)$.

\subsection{Harmonic limit of the triaxial rotor.}

We shall now illustrate the use of the concept of contraction of a
Higgs algebra, in connection with the structure of $H$, of
Eq.(\ref{hrotor}), in the case of large angular momentum $I$. In
the standard representation of angular momentum \cite{ref10}, the
Hamiltonian of equation (\ref{hrotor}) can be analyzed by making
use of boson mapping techniques \cite{ref11}.

We shall use the well known Holstein-Primakoff boson mapping to
get the images of the operators of Eq. (\ref{ec}),

\begin{eqnarray}
I_+ & = & \zeta^\dagger ~\sqrt{2 I - \zeta^\dagger \zeta}, \nonumber \\
I_- & = & \sqrt{2 I - \zeta^\dagger \zeta}~\zeta, \nonumber \\
I_3 & = & I- \zeta^\dagger \zeta.
\end{eqnarray}
where $\zeta^\dagger$ ($\zeta$) are boson creation (annihilation)
operators. In the large angular momentum limit, $I>>1$, the
Hamiltonian of Eq. (\ref{hrotor}) reads

\begin{eqnarray}
H ~~   & = & H_{00}+H_{11}+H_{20}+H_{nn}+H', \nonumber \\
H_{00} & = & A_3 ~I(I+1)- \frac 12 (A_1-A_2) \kappa I, \nonumber \\
H_{11} & = & - \frac 12 (A_1-A_2) \kappa (2I-1) ~\zeta^\dagger \zeta, \nonumber \\
H_{20} & = & ~\frac 12 (A_1-A_2) I ( ~\zeta^{\dagger 2}+ \zeta^2), \nonumber \\
H_{nn} & = & ~\frac 12 (A_1-A_2) \kappa ~(\zeta^\dagger \zeta)^2, \nonumber \\
H' ~~  & = & -\frac 1 {8 I} (A_1-A_2)  ( ~\zeta^{\dagger 3} \zeta+~\zeta^\dagger \zeta^3).
\label{rpa0}
\end{eqnarray}
The terms of $H$ of Eq. (\ref{rpa0}) are classified in powers of
$I$, and by the corresponding powers of the boson creation and
annihilation operators $\zeta^\dagger$ and $\zeta$
\begin{equation}
\sum_{k n} a_{kn}(I)~{\zeta^{\dagger k}} {\zeta^{}}^n.
\end{equation}
The leading order terms, $H_{\rm{RPA}}=H_{00}+H_{11}+H_{20}$, can
be diagonalized by adopting the Random Phase Approximation (RPA)
formalism \cite{ref4,ref11}. This is achieved by  introducing the
RPA boson operator

\begin{eqnarray}
\Gamma^\dagger & = & X~\zeta^\dagger - Y ~\zeta,
\label{phonon}
\end{eqnarray}
and by solving the equation of motion

\begin{eqnarray}
[H_{\rm RPA},\Gamma^\dagger]  = \omega \Gamma^\dagger.\label{rpac}
\end{eqnarray}
Details of the procedure, which is rather well known, can found in
textbooks \cite{ref4,ref12}. In this approximation $H_{\rm RPA}$ takes the form

\begin{eqnarray}
H_{\rm RPA} =  A_3I(I+1) +\omega ~\left(\Gamma^\dagger \Gamma+\frac 12
\right),
\end{eqnarray}
with
\begin{eqnarray}
\omega =  2 I ~\sqrt {(A_3-A_1)(A_3-A_2)}.
\end{eqnarray}
The expression for the RPA eigenfrequency $\omega$ is obtained by
expressing $H_{\rm{RPA}}$ and $\Gamma^{\dagger}$ in terms of the
operators $\zeta^{\dagger}$ and $\zeta$ and by equaling terms
after performing the commutation (\ref{rpac}).

The remaining terms $H_{nn}$ and $H'$ of the Hamiltonian of Eq.
(\ref{rpa0}), can then be transformed to the RPA boson basis
$(\Gamma^\dagger, ~\Gamma)$. By keeping terms up to $\left
(\Gamma^\dagger \Gamma\right)^2$ and neglecting negative powers of
the angular momentum in the coefficients $a_{kn}(I)$, one gets

\begin{eqnarray}
H  \approx  A_3I(I+1)
& + & \left(\omega + \frac 1 {4 \kappa}(A_1-A_2)\right)~\left(\Gamma^\dagger \Gamma+\frac 12 \right), \nonumber \\
& + & \frac 12  (A_1-A_2) \kappa \left(\Gamma^\dagger \Gamma
\right)^2.
\end{eqnarray}

Proceeding now in a manner analogous to the previous subsection,
we shall show that this Hamiltonian is obtained as a particular
case of the $sl^{(3)}(2,\R)$ algebra. As we shall demonstrate, the
harmonic limit of the  triaxial rotor can alternatively be
obtained through a contraction procedure and next embedded, as a
particular case, within the  $sl^{(3)}(2,\R)$ structure together
with some infinite-dimensional representations. In what follows we
describe explicitly this procedure and, furthermore, we shall show
that this is equivalent to the RPA approximation. It means that,
at leading order, both constructions lead to the same results.


The contraction of the Lie algebra (\ref{ec}) to the harmonic
oscillator algebra $h_4$ is obtained by introducing the new
generators defined by~\cite{cont}: \be
\newi_\pm=\varepsilon I_\pm,\quad \newi_3=-I_3 +\frac {\bf 1}{2 \varepsilon^2},
\label{ja} \ee where $\varepsilon$ is the contraction parameter
and ${\bf 1}$ is the identity operator, which obviously commutes
with $I_\pm$ and $I_3$. By computing the new Lie brackets   and
next applying the limit $\varepsilon\to 0$ we find the defining
relations for $h_4$: \be [\newi_3,\newi_\pm]=\pm \newi_\pm,\quad
[\newi_-,\newi_+]={\bf 1},\quad  [{\bf 1},\, \cdot\,]=0.
\label{jb} \ee Since the eigenvalue of ${\bf 1}$ is always 1 in
all the representations here considered, hereafter   we always fix
${\bf 1}\equiv 1$. In this respect, notice that, in fact, such an
identity operator should   also be introduced in the commutation
relations of $sl^{(3)}(2,\R)$ (\ref{sl3}) multiplying the
parameter $\delta$, but due to the above reason this is not
usually  written explicitly in the literature~\cite{r2,r1,r3}.

The contracted Casimir $\newcas^2$ comes from (\ref{ed}) as the
limit \be
\newcas^2=\lim_{\varepsilon\to 0}\varepsilon^2 \left({\bf I}^2 - \frac{1}{4 \varepsilon^4}
\right)=-\newi_3+\frac 12 (\newi_+ \newi_- +\newi_-\newi_+) .
\label{jc} \ee The contraction of the rotor Hamiltonian
(\ref{hrotor}) follows as \bea \hat H &=& \lim_{\varepsilon\to
0}\varepsilon^2 \left(H  - \frac{A_3}{4 \varepsilon^4} \right) \cr
&=& \frac 12 (A_1+A_2) \newcas^2
 +\frac 14 (A_1 - A_2) (\newi_+^2 + \newi_-^2)\cr
&&\quad -\frac 12 (A_1-A_2)\kappa \newi_3 , \label{jd} \eea where
$\kappa$ is the same of Eq.~(\ref{ef}). Now all of these results
can be introduced in $sl^{(3)}(2,\R)$ by taking the parameters
(\ref{db}) with $c=0$, \be \alpha= \beta= 0,\quad \gamma= - 8
\cteb^2,\quad \delta=-2\cteb^2 , \label{je} \ee which  implies
that the eigenvalue of $\newcas^2$ is $1/2$. The generators of
$sl^{(3)}(2,\R)$ are then chosen as \be J_\pm =b \newi_\pm^2,\quad
J_3=\frac 12 \newi_3, \label{jf} \ee and fulfil the commutation
relations \be [J_3,J_\pm]=\pm J_\pm,\quad [J_+,J_-]= - 8 b^2 J_3 -
4 b^2 \newcas^2. \label{jg} \ee Hence the harmonic limit of the
rotor Hamiltonian arises as the $sl^{(3)}(2,\R)$-system given by
\bea &&{\cal H}=\frac 12 (A_1+A_2)\newcas^2
 +\frac 1{4b} (A_1 - A_2) (J_+ + J_-)\cr
&&\qquad -  (A_1-A_2)\kappa J_3 , \label{jeje} \eea which compares
with the Hamiltonian of Eq.(\ref{hrotorhiggs}) up to the term
proportional to $J_3$, due to the contraction (\ref{ja}). Except
for an additive constant, Eq.(\ref{jeje}) coincides with the terms
of Eq.(\ref{rpa0}) labelled as $H_{\rm{RPA}}$.

\section{Results and Discussions}

In this section we shall present and discuss the results of our
calculations. We shall begin with the analysis of the rotational
regime by calculating the eigenvalues of the Hamiltonian of Eq.
(\ref{hrotorhiggs}) and then we shall proceed by showing the
results corresponding to the eigenvalues of the Hamiltonian of Eq.
(\ref{jeje}). Since we have already presented the algebraic
details in the previous sections II and III, we shall focus on the
numerical results.

The spectrum of a triaxial rotor has remarkable features
\cite{ref4} like the breaking of degeneracies and the re-ordering
of levels in degenerate pairs. The transition between both regimes
is governed by the asymmetry parameter $\kappa$. These features
are exhibited, too, by the spectrum of eigenvalues of the
Hamiltonian (\ref{hrotorhiggs}). Figure 1 shows the spectrum as a
function of the asymmetry $\kappa$ and for a fixed angular
momentum, $I=10$. Figure 2 shows the low energy region of the
spectrum for $I=4$. As it can be seen from both figures, the just
mentioned features emerge clearly. It means that the choice
(\ref{choi}) of the non-linear algebra $sl^{(3)}(2,\R)$ leads to
the triaxial rotor. This is an expected result, because using the
algebraic procedure of Section III we have shown that the matrix
elements of both Hamiltonians do indeed coincide. Figure 3 shows
the eigenvalues of the Hamiltonian of Eq. (\ref{jeje}) as a
function of the asymmetry $\kappa$. The spectrum is clearly
harmonic for a fixed value of $\kappa$. The dependence of the
frequency $\omega$ upon $\kappa$ and angular momentum $I$ is a
unique feature of the large angular momentum limit of the triaxial
rotor. The domains $|\kappa|<1$, of the rotor, and $|\kappa|>1$,
of the vibrator are not connected.

\section{Conclusions}

In this work we have studied the correspondence between a
polynomial algebra, of the cubic Higgs-type, and the standard
angular momentum and harmonic oscillator algebras. Particularly,
we have shown that: (i) by a suitable choice of the parameters
entering the definition of the cubic algebra, one may obtain the
triaxial rotor, and, (ii) in the limit of large angular momentum,
the vibrational structure emerges as a contraction of the algebra.

We think that these results suggest the existence of direct
physical realizations of the Higgs algebra for the case of
Hamiltonians describing nuclear vibrations and rotations, as it
was found for the case of classical Hamiltonians, e.g. the study
of the Kepler potential on a two-dimensional curved space.

For the specific application to nuclear structure problems, the
potential of non-linear algebras in dealing with the construction
of Hamiltonians becomes evident. The use of non-linear algebras,
like the $sl^{(3)}(2,\R)$ or even algebras with a larger number of
generators, may shead light on more involved structures lying in
between the rotational and vibrational extremes. This may be the
case of the newly discovered symmetry reported in \cite{casten}.
Work is in progress concerning this point.

\section*{Acknowledgements}

This work was partially supported  by the CONICET (Argentina,
Project PIP-02037), by the Ministerio de Educaci\'on y Ciencia
(Spain, Project FIS2004-07913), and by the Junta de Castilla y
Le\'on   (Spain, Project  VA013C05). A.B. and F.J.H. acknowledge
the hospitality received at the Department of Physics of the
University of La Plata, Argentina.



\newpage

\begin{figure}
\includegraphics[height=8cm,width=6cm]{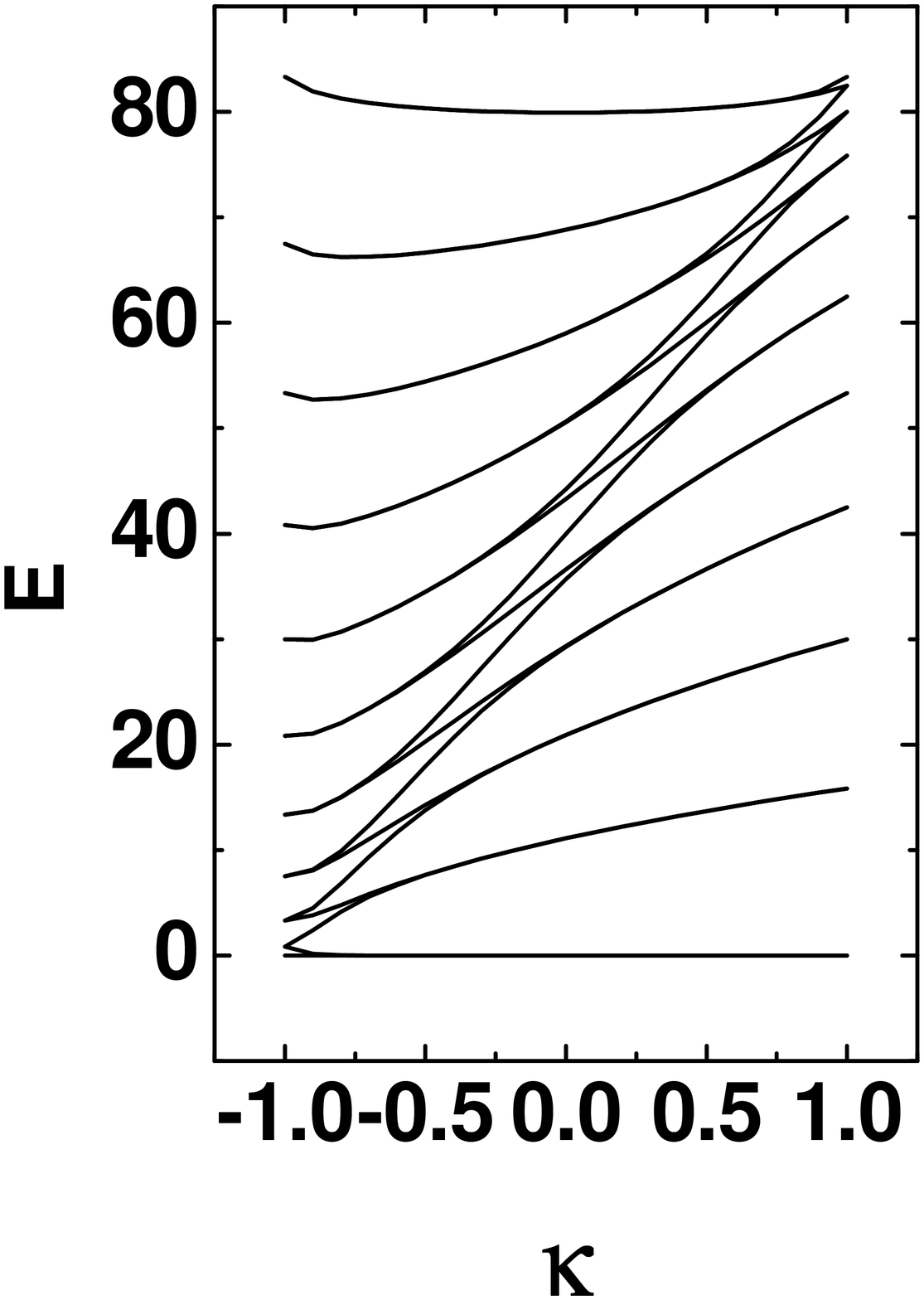}
\caption{Eigenvalues of the Hamiltonian of Eq.
(\ref{hrotorhiggs}), as a function of the asymmetry $\kappa$. The
values are shown in arbitrary units. The parameters used in the
calculations are: $I=10$, ${A}_1=1.66$, and ${A}_2=0.83$. The
results coincide with the ones of a triaxial rotor \cite{ref4}
\label{fig:fig1}}.
\end{figure}

\begin{figure}
\includegraphics[height=8cm,width=6cm]{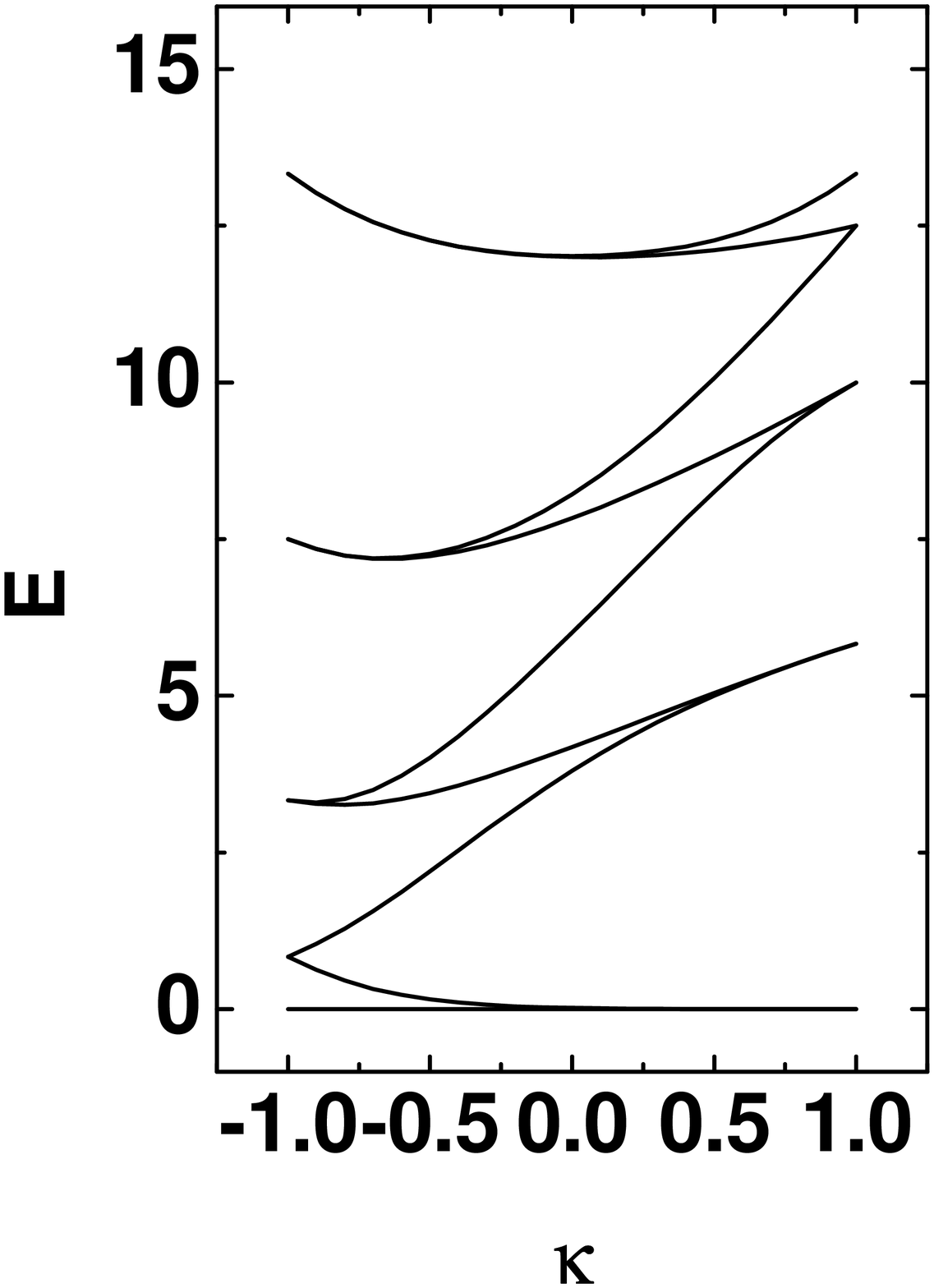}
\caption{Idem as Figure 1, for $I=4$.\label{fig:fig2}}.
\end{figure}

\begin{figure}
\includegraphics[height=8cm,width=6cm]{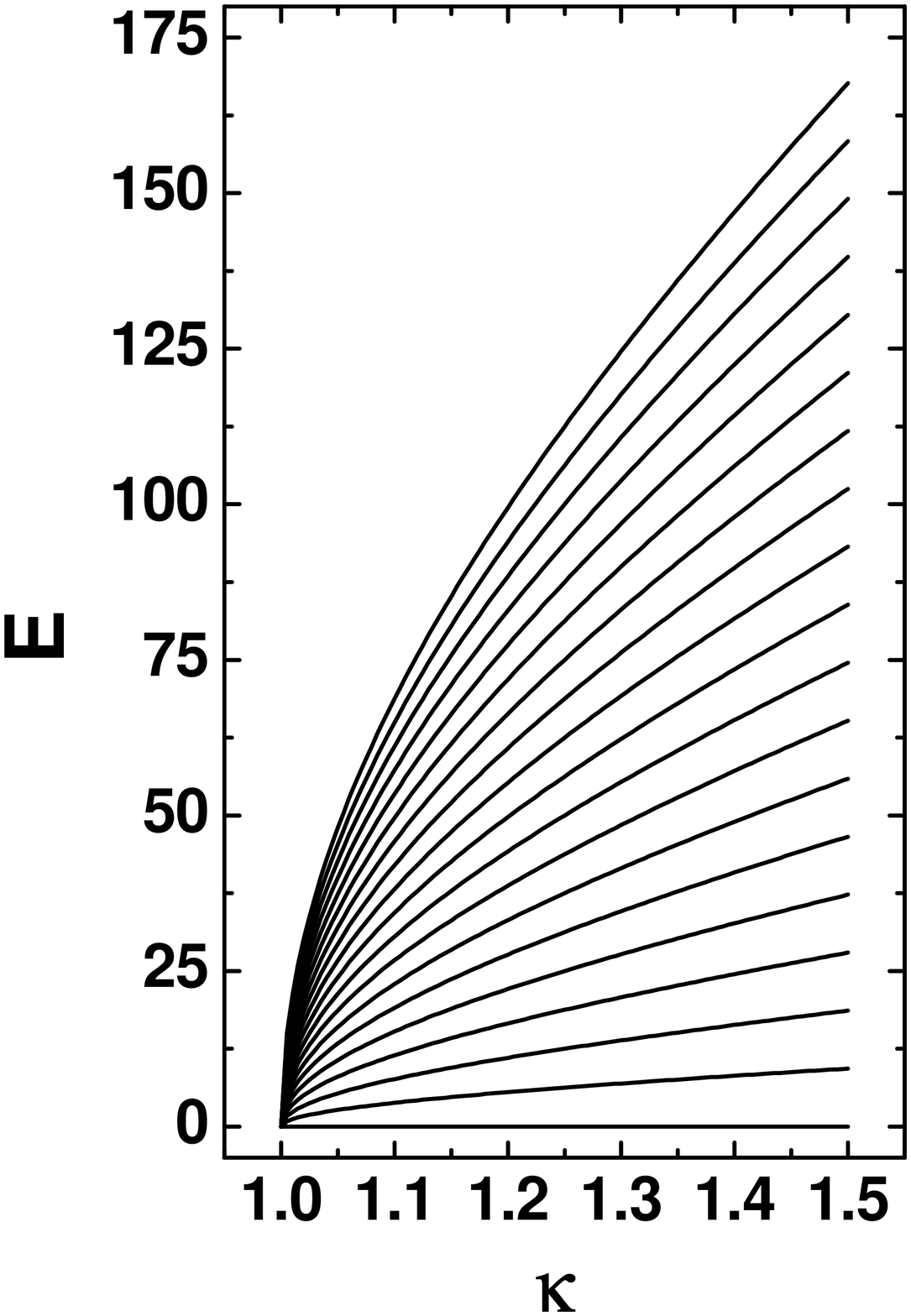}
\caption{Eigenvalues of the Hamiltonian of Eq.(\ref{jeje}), as a
function of the asymmetry $\kappa$. The parameters are the same
given in the captions to Figure 1. In curves are shown the results
of the harmonic limit of Eq. (\ref{hrotorhiggs}) obtained by
contracting the $sl^{(3)}(2,\R)$ algebra.  \label{fig:fig3}}.
\end{figure}


\begin{references}
%
\bibitem {r2} P. W. Higgs,  J. Phys. A {\bf 12}, 309 (1979).
%
\bibitem {r1} N. Debergh, J. Phys. A {\bf 33}, 7109 (2000).
%
\bibitem {r3} N. Debergh, J. Ndimubandi, and B. Van den Bossche,
Mod. Phys. Lett. A {\bf 18}, 1013 (2003).
%
\bibitem{ref4} A. Bohr and B. Mottelson, Nuclear Structure vol. 2
Benjamin, Reading, MA, 1975)
%
\bibitem{ref5} F. Iachello and A. Arima, The interacting boson model. Cambridge Monographs on Mathematical Physics.
Cambridge (2006).
%
\bibitem{ref6} N. Debergh and Fl. Stancu, J. Phys. A {\bf 34}, 3265 (2001).
%
\bibitem {ref6b} V. S. Kumar, B. A. Bambah and R. Jagannathan,
Mod. Phys. Lett. A {\bf 17}, 1559 (2002).
%
\bibitem {ref6c} D. Ruan, Phys. Lett. A {\bf 319}, 122 (2003).
%
\bibitem{ref7} N. Debergh and B. Van den Bossche,
Ann. Phys.  {\bf 308}, 605 (2003).
%

\bibitem{tmp} A. Ballesteros, O. Civitarese, F. J. Herranz and
M. Reboiro, Theor. Math. Phys. {\bf 137}, 1495 (2003).

%
\bibitem{ref10} D. M.Brink and G. R. Satchler, Angular Momentum.
Oxford Library of Physical Sciences Series, Oxford, 1994
%
\bibitem{ref11} A. Klein and E. R. Marshalek,
Rev. of Modern Phys. {\bf 63}, 375(1991).
%
\bibitem{ref12} P. Ring and P. Shuck, The nuclear many-body problem.
(Springer Verlag, N.Y. 1980)
%
\bibitem{cont} A. Ballesteros, F. J. Herranz and P. Parashar, J. Phys. A {\bf 32}, 2369 (1999)
\bibitem{casten} E.A. McCutchan and R. F. Casten, Phys. Rev. C {\bf
74}, 057302(2006); E.A. McCutchan, N. V. Zamfir and R. F. Casten,
Phys. Rev. C {\bf 71}, 034309 (2005); E.A. McCutchan et al. ,
Phys. Rev. C {\bf 71}, 024309(2006).

\end{references}
\end{document}